\documentclass[prl,aps,preprint,superscriptaddress,showpacs]{revtex4}
\usepackage{epsfig,amsmath}
\usepackage{bm}
\usepackage{longtable}

\DeclareMathOperator{\Sp}{Sp}

\begin{document}

\title{Symmetries of the Bell correlation inequalities}

\author{Cezary \'Sliwa}
\affiliation{Centre for Theoretical Physics,
Al.\ Lotnik\'ow 32/46, 02-668 Warsaw, Poland}

\date{\today}

\begin{abstract}
  Bell correlation inequalities for two sites and $2+n$ or $3+3$
  two-way measurements (``dichotomic observables'') are considered.
  In the $2+n$ case, any facet of the classical experience polytope is
  defined by a CHSH inequality involving only two pairs of the
  observables. In the $3+3$ case, contrary to earlier results, the
  action of the symmetry group reduces the set of all Bell
  inequalities to just 3 orbits, only one of them being ``new'' (not
  known from the $2+2$ case). A detailed calculation for the singlet
  state of two qubits reveals the configurations of a maximal
  violation for this class of inequalities.
\end{abstract}


\maketitle

Although the notion of a general \emph{Bell inequality} seems to be
well established by now \cite{WW2001}, examples that can be found
in the present literature rarely go beyond the well-known
Clauser-Horne-Shimony-Holt (CHSH) inequality,
\begin{equation}
  \label{eq:chsh}
  \frac12 [E(A_1 B_1) + E(A_1 B_2) + E(A_2 B_1) - E(A_2 B_2)] \leq 1.
\end{equation}
It is so because, despite the great simplicity of the basic idea,
any attempt to find effectively a complete set of Bell inequalities
for a given set of observables encounters computational difficulties
that can be overcome only for a few simple cases. Two such cases
has been considered in~\cite{PS2001}, namely the case of two observables
at each of three sites and three observables at each of two sites,
with observables having two possible values. On the other hand,
depending on the configuration of the observables,
the problem admits a number of symmetries, like permuting
observables, sites with identical configuration
of observables or values of a given observable.
It is not clear whether this symmetries can be employed
to reduce the complexity of the convex-hull problem.
However, what one wants to find, are the \emph{classes}
of equivalent facets of the correlation polytope,
rather than all the facets themselves. Therefore,
following a convex hull computation,
one should split the set of all Bell inequalities
into orbits of the symmetry group, and take just one
representative from each orbit.

We can assume the two values of the observables to be $\{ +1, -1 \}$.
Under this assumption the notation $(2, 2)$ and $(2, 2, 2)$ or $(3, 3)$
unambiguously describes the case of~(\ref{eq:chsh})
and the two cases considered in~\cite{PS2001}.
Let us label the sites~$X$ with letters $A, B, C$, and denote the
observables respectively by $A_i, B_j, C_k$. With this
notation, the symmetries are: $X_i \to -X_i$, $X_i \to X_{\sigma(i)}$,
$X_i \to \sigma(X)_i$, where the $\sigma$ denotes a permutation.
To illustrate a method of systematic treatment
of the no-signaling constraints, instead of
the unknown probability distribution on the space of all possible
classical configurations~$\Lambda$
we consider as coordinates all the correlations of the form
\begin{equation}
  \label{eq:co}
  E_{i_A i_B \ldots } = E(A_{i_A}^{n_A-1} B_{i_B}^{n_B-1} \ldots), \quad
  n_X = 1, 2, \ldots, |\Sp(X_{i_X})|.
\end{equation}
For the simplest cases considered in~\cite{PS2001},
the coordinates are the expected values
$E(X_i)$, $E(X_i Y_j)$, $E(X_i Y_j Z_k)$, where $X_i Y_j$
denotes the product of the two (fictitious) classical random
variables that we associate with the outcome of (independent)
measurements at the sites $X$ and~$Y$.
Now, Bell inequalities correspond to the facets of the (purely
classical) \emph{correlation polytope} that is obtained
by projecting the simplex of classical probability
distributions on~$\Lambda$ onto the coordinates
introduced by~(\ref{eq:co}).
This procedure respects the no-signaling constraints
and the uncertainty principle that let us measure simultaneously
only one observable at each site, and preserves
all the information available from a measurement
(further projection, e.g.\ taking into account only
full correlations \cite[section 3.2]{WW2001},
leads to other types of Bell inequalities).
In short, the vertices of the correlation polytope are obtained
by enumerating the elements $\lambda \in \Lambda$,
and calculating for each~$\lambda$ all the products
that appear in~(\ref{eq:co}).

The case $(2, 2)$ has been discussed already by~\cite{F1982}.
Any Bell inequality is of the form (\ref{eq:chsh}).
The extremality of this inequality manifests through
a feature that is usually silently ignored: the expression
\begin{equation}
  f_{2}(a_1, a_2; b_1, b_2) =
    \frac12 (a_1 b_1 + a_1 b_2 + a_2 b_1 - a_2 b_2)
\end{equation}
takes \emph{only} the extremal values $\{ +1, -1 \}$
if the arguments do so. Hence, it may be considered
a \emph{boolean function} of the four arguments.
In this approach, instead of checking that the values
are in the required range, one checks that
the expected value of $f_{2}$ is well-defined in quantum mechanics,
i.e.\ that $f_{2}$ is a sum of terms none of which
depends on both $a_1$ and~$a_2$, or both $b_1$ and $b_2$.
Moreover, up to the symmetries $a_1 \leftrightarrow a_2$,
$b_1 \leftrightarrow b_2$, $a_i \rightarrow -a_i$,
$b_i \rightarrow -b_i$, $(a_1, a_2) \leftrightarrow (b_1, b_2)$,
it is the \emph{only} boolean function
with these properties that non-trivially depends
on all the four arguments. Other such functions
can be obtained by substitutions, like $a_1 \rightarrow +1$,
$b_2 \rightarrow b_1$, etc.
Let us mention that this includes any boolean function
of two arguments, like $f(a_1; b_1)$, in the form
$f(a; b) = \pm f_{2}(a, \pm 1; b, \pm 1)$,
$f(a; b) = \pm f_{2}(a, +1; b, b)$.

The results obtained for the $(3, 3)$ case are definitely
interesting. Among all the $684$ inequalities,
there are $576$ of the form
\begin{eqnarray}
  \label{eq:b3}
  0 & \leq & 4 + E(A_1) + E(A_2) + E(B_1) + E(B_2) \\
  & & {} + E(A_1 B_1) + E(A_1 B_2) + E(A_2 B_1) + E(A_2 B_2) \nonumber \\
  & & {} + E(A_3 B_1) - E(A_3 B_2) + E(A_1 B_3) - E(A_2 B_3), \nonumber
\end{eqnarray}
$72$ are the variations of the CHSH ineq.\ (\ref{eq:chsh})
corresponding to choosing two observables at each site,
and the remaining $36$ guarantee that the probability distribution
for the result of any joint measurement is positive.
One can easily check the counts to convince oneself that, as expected,
the orbits of the symmetry group action are full.
A convenient way to handle the symmetries is to organize
the coordinates into a rank-2 tensor
$[E(\tilde A_i, \tilde B_j)]_{ij}$,
with $(\tilde X_i) = (1, X_1, X_2, X_3)$. Then,
the $3+3$ Bell inequality takes the form
\begin{equation}
  0 \leq \alpha_{ij} E(\tilde A_i, \tilde B_j), \quad
  (\alpha_{ij}) = \left( 
    \begin{array}{rrrr}
      4& 1& 1& 0\\ 1& 1& 1& 1\\ 1& 1& 1& -1\\ 0& 1& -1& 0
    \end{array}
  \right).
\end{equation}

Certainly, it is the possibility of a quantum violation
that makes the ineq.\ (\ref{eq:b3}) a true Bell inequality.
It is quite easy to demonstrate this possibility for
a singlet state of two qubits,
$\left| \psi \right> =
  \frac{1}{\sqrt{2}} ( \left| \uparrow \downarrow \right>
  - \left| \downarrow \uparrow \right>$.
The well-known relations
for Pauli matrices let us write any observable~$\hat A \in B(H_2)$,
$\hat A^2 = 1$ in the form
$\hat A = {\bf a} \cdot \hat {\bm{\sigma}} =
  a_x \hat \sigma_x + a_y \hat \sigma_y + a_z \hat \sigma_z$,
where ${\bf a}$ is a unit vector,
$\left\|{\bf a}\right\|^2 = a_x^2 + a_y^2 + a_z^2 = 1$.
The R.H.S of~(\ref{eq:b3}) is equal to the expected value of
$\hat F ( {\bf a}_i, {\bf b}_i ) =
  4 + ( \hat A_1 + \hat A_2 ) + ( \hat B_1 + \hat B_2 )
  + ( \hat A_1 + \hat A_2 ) ( \hat B_1 + \hat B_2 )
  + \hat A_3 ( \hat B_1 - \hat B_2 ) + ( \hat A_1 - \hat A_2 ) \hat B_3$,
where the unit vectors ${\bf a}_i$, ${\bf b}_j$ have been
introduced to describe what observables $\hat A_i$, $\hat B_j$
are to be measured at each site. In terms of ${\bf a}_i$
and ${\bf b}_j$, the measured correlations are
$\left< \psi \right| \hat A_i \left| \psi \right> = 0$,
$\left< \psi \right| \hat B_j \left| \psi \right> = 0$, and
$\left< \psi \right| \hat A_i \hat B_j \left| \psi \right> =
  -  {\bf a}_i \cdot {\bf b}_j$, thus
\begin{equation}
  \left< \psi \right| \hat F ( {\bf a}_i, {\bf b}_i ) \left| \psi \right> =
  4 - {\bf a}_3 \cdot ( {\bf b}_1 - {\bf b}_2 )
  - {\bf a}_1 \cdot ( {\bf b}_1 + {\bf b}_2 + {\bf b}_3 )
  - {\bf a}_2 \cdot ( {\bf b}_1 + {\bf b}_2 - {\bf b}_3 ).
\end{equation}
For given ${\bf b}_j$'s, the extremal value,
\begin{equation}
  4 - \left\| {\bf b}_1 - {\bf b}_2 \right\|
  - \left\| {\bf b}_1 + {\bf b}_2 + {\bf b}_3 \right\|
  - \left\| {\bf b}_1 + {\bf b}_2 - {\bf b}_3 \right\|,
\end{equation}
is attained when
${\bf a}_1 \propto ( {\bf b}_1 + {\bf b}_2 + {\bf b}_3 )$, etc.
Fortunately, $({\bf b}_1 + {\bf b}_2)$ and $({\bf b}_1 - {\bf b}_2)$
are orthogonal. Let $\beta$ be the angle between
$( {\bf b}_1 + {\bf b}_2 )$ and ${\bf b}_3$. Hence,
$\left\| {\bf b}_1 - {\bf b}_2 \right\| = 2 \cos\alpha$,
$\left\| {\bf b}_1 + {\bf b}_2 \right\| = 2 \sin\alpha$,
$\left\| {\bf b}_1 + {\bf b}_2 \pm {\bf b}_3 \right\| =
  \sqrt{(2\sin\alpha \pm \cos\beta)^2 + \sin^2\beta}$.
We obtain the extremal violation (by $-1$) for
$\alpha=\pi/3$, $\beta=\pi/2$, but there is still
one degree of freedom in choosing the extremal configuration:
the angle between $( {\bf b}_1 - {\bf b}_2 )$ and ${\bf b}_3$.
In the spherical coordinates,
${\bf a} = (a_x, a_y, a_z) = ( \sin \theta_a \cos \phi_a,
  \sin \theta_a \sin \phi_a, \cos \theta_a )$,
the extremal configurations are
(up to a rotation/unitary transform in~$H_2$):
$\theta_{a1} = \theta_{a2} = \theta_{b1} = \theta_{b2} = \pi/6$,
$\theta_{a3} = \theta_{b3} = \pi/2$,
$\phi_{a1} = \phi_{a2} + \pi = \phi_{b3}$, 
$\phi_{b1} = \phi_{b2} + \pi = \phi_{a3}$.

Now that we have a Bell inequality for $3+3$ observables,
one asks the question what we gain using it instead
of~(\ref{eq:chsh}). Comparing the raw numbers of maximum violation
is certainly meaningless, unless some normalization is defined.
One can normalize, e.g., the free numeric coefficient
or the classical range of the inequality's R.H.S.,
$E = \sum_{ij} \alpha_{ij} E_{ij}$.
The natural choice is, however, to normalize the standard deviation
\begin{equation}
  \Delta E = \sqrt{\sum\nolimits_{ij} \alpha_{ij} (\Delta E_{ij})^2},
\end{equation}
where $\Delta E_{ij}$ is the standard deviation
of the measurement outcome for $E(\tilde A_i, \tilde B_j)$.
In the simplest case, each $E_{ij}$ is determined by
a separate sequence of measurements,
with an equal number of individual measurements for each
configuration~$(i, j)$. In this case,
\begin{equation}
  (\Delta E_{ij})^2 \propto
  \left[
  \left< \psi \right| \left( \hat{\tilde A}_i \hat{\tilde B}_j \right)^2
    \left| \psi \right>
  - \left( \left< \psi \right| \hat{\tilde A}_i \hat{\tilde B}_j
    \left| \psi \right> \right)^2
  \right]
\end{equation}
(the possible optimization is to derive $E_{1j}$ and $E_{i1}$
from the remaining $E_{ij}$'s, but this would result
in correlations between the components, invalidating the above formula).
The maximal ratio $|E/\Delta E|$ is (proportional to)
$0.585786$ for~(\ref{eq:chsh}), and $0.342997$ for~(\ref{eq:b3}).
Therefore, for the singlet state, the well-known CHSH-inequality
is not only significantly simpler, but also stronger than~(\ref{eq:b3}).
Nevertheless, since a complete set of Bell inequalities
for $3+3$ observables includes~(\ref{eq:b3}), there may exist states
that satisfy~(\ref{eq:chsh}), even for any $\hat A_i$, $\hat B_j$,
but violate~(\ref{eq:b3}).

Let us now consider the $(2, n)$ case. This corresponds
to a $d$-dimensional convex-hull problem, with $d=3n+2$.
Any facet of the correlation polytope coincides with
$d$ vertices in general position, i.e.\ the vectors
from a chosen reference vertex to the remaining vertices must span
a subspace of dimension $d-1$. Let $\alpha_{ij}$
be the coefficients of the corresponding inequality.
Alternatively, one can classify the vertices by the values
$(a_1, a_2)$ of $(A_1, A_2)$,
and choose one reference vertex in each of the classes.
Then, if there are two vertices in the class $(a_1, a_2)$
that differ in $b_j$,
$\alpha_{0j} + a_1 \alpha_{1j} + a_2 \alpha_{2j} = 0$.
The number of such $j$'s in a class is the dimension
of the subspace spanned by the corresponding vectors.
The sum of this numbers must be at least $d-4$,
as there is one reference vertex in each class
and there can be at most $4$ classes. Hence,
since $3n-2 > 2n$ for $n > 2$,
for some $j$ there are $3$ such equations. Consequently,
$\alpha_{0j} = \alpha_{1j} = \alpha_{2j} = 0$,
and $B_j$ does not appear in the corresponding inequality.
By induction, any Bell inequality for $2+n$ variables
$(A_1, A_2, B_1, B_2, \ldots, B_n)$ involves only
one pair of~$B_j$'s, and therefore is of the CHSH-form.

A slightly more general argument can be applied to the case
$(k, n)$, with $k > 2$ observables at one site to show
that there are no Bell inequalities beyond the limit $n \leq  2^k-2$.
Since this upper bound depends exponentially on~$k$, its practical
significance may probably be limited.

Similar results have been obtained for other cases.
In particular, for the case $(2, 2, 2)$ of three pairs of observables,
there are $46$ classes of inequalities. They are listed
in Table~\ref{tab1}. Although it is possible 
a to reconstruct a number of them by chaining
the boolean function~$f_2$, the interpretation
and structure of those inequalities remain mostly unknown.

In summary, a method of systematic treatment of the Bell
correlation inequalities and their symmetries has been proposed.
Detailed results for several special cases have been presented,
including some combinatorial properties of the CHSH-inequality.

\clearpage

\begin{longtable}{|c|c|}
\caption{Bell inequalities for three pairs of observables}\label{tab1}\\
\hline
Class& Representative\\
\endfirsthead
\hline
Class& Representative\\
\endhead
\hline
   1 &  \parbox[t]{5in}{\raggedright $0 \leq 1 - E(A_1) - E(B_1) + E(A_1 B_1) - E(C_1) + E(A_1 C_1) + E(B_1 C_1) - E(A_1 B_1 C_1)$\\\vspace{6pt}} \\ \hline
   2 &  \parbox[t]{5in}{\raggedright $0 \leq 2 - E(A_1 B_1 C_1) - E(A_2 B_2 C_1) - E(A_2 B_1 C_2) + E(A_1 B_2 C_2)$\\\vspace{6pt}} \\ \hline
   3 &  \parbox[t]{5in}{\raggedright $0 \leq 2 - E(A_1 B_1 C_1) - E(A_2 B_1 C_1) - E(A_1 B_2 C_2) + E(A_2 B_2 C_2)$\\\vspace{6pt}} \\ \hline
   4 &  \parbox[t]{5in}{\raggedright $0 \leq 2 - 2 E(A_1) - E(B_1 C_1) + E(A_1 B_1 C_1) - E(B_2 C_1) + E(A_1 B_2 C_1) - E(B_1 C_2) + E(A_1 B_1 C_2) + E(B_2 C_2) - E(A_1 B_2 C_2)$\\\vspace{6pt}} \\ \hline
   5 &  \parbox[t]{5in}{\raggedright $0 \leq 3 - E(A_1) - E(B_1) - E(A_2 B_1) - E(A_1 B_2) + E(A_2 B_2) - E(C_1) - E(A_2 C_1) + E(A_1 B_1 C_1) + E(A_2 B_1 C_1) - E(B_2 C_1) + E(A_1 B_2 C_1) - E(A_1 C_2) + E(A_2 C_2) - E(B_1 C_2) + E(A_1 B_1 C_2) + E(B_2 C_2) - E(A_2 B_2 C_2)$\\\vspace{6pt}} \\ \hline
   6 &  \parbox[t]{5in}{\raggedright $0 \leq 3 - E(A_1) - E(B_1) - E(A_1 B_1) - E(C_1) - E(A_2 C_1) + E(A_1 B_1 C_1) + E(A_2 B_1 C_1) - E(B_2 C_1) + E(A_1 B_2 C_1) - E(A_1 C_2) + E(A_2 C_2) + E(B_1 C_2) - E(A_2 B_1 C_2) - E(B_2 C_2) + E(A_1 B_2 C_2)$\\\vspace{6pt}} \\ \hline
   7 &  \parbox[t]{5in}{\raggedright $0 \leq 4 - 3 E(A_1 B_1 C_1) - E(A_2 B_1 C_1) - E(A_1 B_2 C_1) + E(A_2 B_2 C_1) - E(A_1 B_1 C_2) + E(A_2 B_1 C_2) + E(A_1 B_2 C_2) - E(A_2 B_2 C_2)$\\\vspace{6pt}} \\ \hline
   8 &  \parbox[t]{5in}{\raggedright $0 \leq 4 - E(A_1 B_1) - E(A_2 B_1) - E(A_1 B_2) - E(A_2 B_2) - 2 E(A_1 B_1 C_1) + 2 E(A_2 B_2 C_1) - E(A_1 B_1 C_2) + E(A_2 B_1 C_2) + E(A_1 B_2 C_2) - E(A_2 B_2 C_2)$\\\vspace{6pt}} \\ \hline
   9 &  \parbox[t]{5in}{\raggedright $0 \leq 4 - E(A_1 B_1) - E(A_2 B_1) - E(A_1 B_2) - E(A_2 B_2) - 2 E(A_1 B_1 C_1) + 2 E(A_1 B_2 C_1) - E(A_1 B_1 C_2) + E(A_2 B_1 C_2) - E(A_1 B_2 C_2) + E(A_2 B_2 C_2)$\\\vspace{6pt}} \\ \hline
  10 &  \parbox[t]{5in}{\raggedright $0 \leq 4 - E(A_1 B_1) - E(A_2 B_1) - E(A_1 B_2) - E(A_2 B_2) - E(A_1 C_1) + E(A_2 C_1) - E(B_1 C_1) - E(A_1 B_1 C_1) + E(B_2 C_1) + E(A_2 B_2 C_1) - E(A_1 C_2) + E(A_2 C_2) + E(B_1 C_2) - E(A_2 B_1 C_2) - E(B_2 C_2) + E(A_1 B_2 C_2)$\\\vspace{6pt}} \\ \hline
  11 &  \parbox[t]{5in}{\raggedright $0 \leq 4 - 2 E(A_1 B_1) - 2 E(A_2 B_2) - E(A_1 B_1 C_1) - E(A_2 B_1 C_1) + E(A_1 B_2 C_1) + E(A_2 B_2 C_1) - E(A_1 B_1 C_2) + E(A_2 B_1 C_2) - E(A_1 B_2 C_2) + E(A_2 B_2 C_2)$\\\vspace{6pt}} \\ \hline
  12 &  \parbox[t]{5in}{\raggedright $0 \leq 4 - 2 E(A_1 B_1) - 2 E(A_2 B_2) - E(A_1 C_1) - E(A_2 C_1) + E(B_1 C_1) - E(A_2 B_1 C_1) + E(B_2 C_1) + E(A_1 B_2 C_1) - E(A_1 C_2) - E(A_2 C_2) + E(B_1 C_2) + E(A_2 B_1 C_2) + E(B_2 C_2) - E(A_1 B_2 C_2)$\\\vspace{6pt}} \\ \hline
  13 &  \parbox[t]{5in}{\raggedright $0 \leq 4 - 2 E(A_1 B_1) - 2 E(A_2 B_1) - E(A_1 B_1 C_1) + E(A_2 B_1 C_1) - E(A_1 B_2 C_1) + E(A_2 B_2 C_1) - E(A_1 B_1 C_2) + E(A_2 B_1 C_2) + E(A_1 B_2 C_2) - E(A_2 B_2 C_2)$\\\vspace{6pt}} \\ \hline
  14 &  \parbox[t]{5in}{\raggedright $0 \leq 4 - 2 E(A_1 B_1) - 2 E(A_2 B_1) - E(A_1 C_1) + E(A_2 C_1) - E(A_1 B_2 C_1) + E(A_2 B_2 C_1) - E(A_1 C_2) + E(A_2 C_2) + E(A_1 B_2 C_2) - E(A_2 B_2 C_2)$\\\vspace{6pt}} \\ \hline
  15 &  \parbox[t]{5in}{\raggedright $0 \leq 4 - 2 E(A_1 B_1) - 2 E(A_2 B_1) - E(A_1 C_1) - E(A_2 C_1) + 2 E(B_1 C_1) - E(A_1 B_2 C_1) + E(A_2 B_2 C_1) - E(A_1 C_2) - E(A_2 C_2) + 2 E(B_1 C_2) + E(A_1 B_2 C_2) - E(A_2 B_2 C_2)$\\\vspace{6pt}} \\ \hline
  16 &  \parbox[t]{5in}{\raggedright $0 \leq 4 - E(A_1) - E(A_2) - E(A_1 B_1) - E(A_2 B_1) - E(A_1 C_1) - E(A_2 C_1) + 2 E(A_2 B_1 C_1) - E(A_1 B_2 C_1) + E(A_2 B_2 C_1) - E(A_1 B_1 C_2) + E(A_2 B_1 C_2) + E(A_1 B_2 C_2) - E(A_2 B_2 C_2)$\\\vspace{6pt}} \\ \hline
  17 &  \parbox[t]{5in}{\raggedright $0 \leq 4 - E(A_1) - E(A_2) - E(A_1 B_1) - E(A_2 B_1) - E(A_1 C_1) - E(A_2 C_1) + E(A_1 B_1 C_1) + E(A_2 B_1 C_1) - 2 E(A_1 B_2 C_2) + 2 E(A_2 B_2 C_2)$\\\vspace{6pt}} \\ \hline
  18 &  \parbox[t]{5in}{\raggedright $0 \leq 4 - E(A_1) - E(A_2) - E(A_1 B_1) - E(A_2 B_1) - E(A_1 C_1) - E(A_2 C_1) + 2 E(B_1 C_1) - E(A_1 B_2 C_1) + E(A_2 B_2 C_1) - E(A_1 B_1 C_2) + E(A_2 B_1 C_2) - 2 E(B_2 C_2) + E(A_1 B_2 C_2) + E(A_2 B_2 C_2)$\\\vspace{6pt}} \\ \hline
  19 &  \parbox[t]{5in}{\raggedright $0 \leq 4 - E(A_1) - E(A_2) - E(A_1 B_1) - E(A_2 B_1) - E(A_1 C_1) - E(A_2 C_1) + 2 E(B_1 C_1) - 2 E(B_2 C_1) + E(A_1 B_2 C_1) + E(A_2 B_2 C_1) - E(A_1 B_1 C_2) + E(A_2 B_1 C_2) - E(A_1 B_2 C_2) + E(A_2 B_2 C_2)$\\\vspace{6pt}} \\ \hline
  20 &  \parbox[t]{5in}{\raggedright $0 \leq 4 - E(A_1) - E(A_2) - E(A_1 B_1) + E(A_2 B_1) - E(A_1 B_2) + E(A_2 B_2) - E(A_1 C_1) + E(A_2 C_1) + E(B_1 C_1) - E(A_1 B_1 C_1) - E(A_2 B_1 C_1) + E(B_2 C_1) - E(A_1 B_2 C_1) - E(A_2 B_2 C_1) - E(B_1 C_2) + E(A_1 B_1 C_2) + E(A_2 B_1 C_2) + E(B_2 C_2) - E(A_1 B_2 C_2) - E(A_2 B_2 C_2)$\\\vspace{6pt}} \\ \hline
  21 &  \parbox[t]{5in}{\raggedright $0 \leq 4 - E(A_1) - E(A_2) - E(B_1) - E(A_1 B_1) - E(B_2) + E(A_2 B_2) - E(A_1 C_1) - E(A_2 C_1) - E(B_1 C_1) + 2 E(A_1 B_1 C_1) + E(A_2 B_1 C_1) - E(B_2 C_1) + E(A_1 B_2 C_1) - E(A_1 B_1 C_2) + E(A_2 B_1 C_2) + E(A_1 B_2 C_2) - E(A_2 B_2 C_2)$\\\vspace{6pt}} \\ \hline
  22 &  \parbox[t]{5in}{\raggedright $0 \leq 4 - E(A_1) - E(A_2) - E(B_1) - E(A_1 B_1) - E(B_2) + E(A_2 B_2) - E(C_1) - E(A_1 C_1) - E(B_1 C_1) + 2 E(A_1 B_1 C_1) + E(A_2 B_1 C_1) + E(A_1 B_2 C_1) - E(A_2 B_2 C_1) - E(C_2) + E(A_2 C_2) + E(A_1 B_1 C_2) - E(A_2 B_1 C_2) + E(B_2 C_2) - E(A_1 B_2 C_2)$\\\vspace{6pt}} \\ \hline
  23 &  \parbox[t]{5in}{\raggedright $0 \leq 4 - E(A_1) - E(A_2) - E(B_1) + E(A_1 B_1) + E(A_2 B_1) - E(B_2) + E(A_1 B_2) + E(A_2 B_2) - E(A_1 C_1) + E(A_2 C_1) + E(A_1 B_1 C_1) - E(A_2 B_1 C_1) + E(A_1 B_2 C_1) - E(A_2 B_2 C_1) - E(B_1 C_2) + E(A_1 B_1 C_2) + E(A_2 B_1 C_2) + E(B_2 C_2) - E(A_1 B_2 C_2) - E(A_2 B_2 C_2)$\\\vspace{6pt}} \\ \hline
  24 &  \parbox[t]{5in}{\raggedright $0 \leq 5 - E(A_1) - E(B_1) - E(A_2 B_1) - E(A_1 B_2) - E(A_2 B_2) - E(C_1) - E(A_2 C_1) + E(B_1 C_1) - 2 E(A_1 B_1 C_1) + E(A_2 B_1 C_1) + 2 E(A_2 B_2 C_1) - E(A_1 C_2) - E(A_2 C_2) + 2 E(A_2 B_1 C_2) + E(A_1 B_2 C_2) - E(A_2 B_2 C_2)$\\\vspace{6pt}} \\ \hline
  25 &  \parbox[t]{5in}{\raggedright $0 \leq 5 - E(A_1) - E(B_1) - E(A_2 B_1) - E(A_1 B_2) - E(A_2 B_2) - E(C_1) - E(A_2 C_1) + E(B_1 C_1) - 2 E(A_1 B_1 C_1) + E(A_2 B_1 C_1) + 2 E(A_2 B_2 C_1) - E(A_1 C_2) - E(A_2 C_2) + 2 E(A_1 B_1 C_2) - E(A_1 B_2 C_2) + E(A_2 B_2 C_2)$\\\vspace{6pt}} \\ \hline
  26 &  \parbox[t]{5in}{\raggedright $0 \leq 5 - E(A_1) - E(B_1) - E(A_1 B_1) - 2 E(A_2 B_2) - E(C_1) - E(A_1 C_1) - E(B_1 C_1) + E(A_1 B_1 C_1) + 2 E(A_2 B_2 C_1) - 2 E(A_2 C_2) + 2 E(A_2 B_1 C_2) + 2 E(B_2 C_2) - 2 E(A_1 B_2 C_2)$\\\vspace{6pt}} \\ \hline
  27 &  \parbox[t]{5in}{\raggedright $0 \leq 5 - 2 E(A_1) - E(A_2) - E(B_1) + E(A_1 B_1) - E(A_1 B_2) - E(A_2 B_2) - E(C_1) + E(A_1 C_1) - 2 E(A_1 B_1 C_1) + 2 E(A_2 B_1 C_1) - E(B_2 C_1) + E(A_1 B_2 C_1) - E(A_1 C_2) - E(A_2 C_2) - E(B_1 C_2) + E(A_1 B_1 C_2) - E(B_2 C_2) + 2 E(A_1 B_2 C_2) + E(A_2 B_2 C_2)$\\\vspace{6pt}} \\ \hline
  28 &  \parbox[t]{5in}{\raggedright $0 \leq 6 - E(A_1) - E(A_2) - E(A_1 B_1) + E(A_2 B_1) - E(A_1 C_1) + E(A_2 C_1) + E(B_1 C_1) - 2 E(A_1 B_1 C_1) - E(A_2 B_1 C_1) - E(B_2 C_1) + E(A_1 B_2 C_1) + 2 E(A_2 B_2 C_1) - E(B_1 C_2) + E(A_1 B_1 C_2) + 2 E(A_2 B_1 C_2) - E(B_2 C_2) + 3 E(A_1 B_2 C_2)$\\\vspace{6pt}} \\ \hline
  29 &  \parbox[t]{5in}{\raggedright $0 \leq 6 - E(A_1) - E(A_2) - E(A_1 B_1) + E(A_2 B_1) - E(A_1 C_1) + E(A_2 C_1) + E(B_1 C_1) - 2 E(A_1 B_1 C_1) - E(A_2 B_1 C_1) - E(B_2 C_1) + E(A_1 B_2 C_1) + 2 E(A_2 B_2 C_1) - E(B_1 C_2) + 3 E(A_1 B_1 C_2) - E(B_2 C_2) + E(A_1 B_2 C_2) + 2 E(A_2 B_2 C_2)$\\\vspace{6pt}} \\ \hline
  30 &  \parbox[t]{5in}{\raggedright $0 \leq 6 - E(A_1) - E(A_2) - 2 E(A_1 B_1) + 2 E(A_2 B_1) - E(A_1 B_2) + E(A_2 B_2) - E(A_1 C_1) + E(A_2 C_1) + E(B_1 C_1) - 2 E(A_1 B_1 C_1) - E(A_2 B_1 C_1) + E(B_2 C_1) - E(A_1 B_2 C_1) - 2 E(A_2 B_2 C_1) - E(B_1 C_2) + 2 E(A_1 B_1 C_2) + E(A_2 B_1 C_2) + E(B_2 C_2) - 2 E(A_1 B_2 C_2) - E(A_2 B_2 C_2)$\\\vspace{6pt}} \\ \hline
  31 &  \parbox[t]{5in}{\raggedright $0 \leq 6 - E(A_1) - E(A_2) - E(B_1) + E(A_2 B_1) - E(B_2) + E(A_1 B_2) - E(A_1 C_1) + E(A_2 C_1) - 2 E(A_2 B_1 C_1) + E(A_1 B_2 C_1) - 3 E(A_2 B_2 C_1) - E(B_1 C_2) + 2 E(A_1 B_1 C_2) + E(A_2 B_1 C_2) + E(B_2 C_2) - 2 E(A_1 B_2 C_2) - E(A_2 B_2 C_2)$\\\vspace{6pt}} \\ \hline
  32 &  \parbox[t]{5in}{\raggedright $0 \leq 6 - E(A_1) - E(A_2) - E(B_1) + E(A_2 B_1) - E(B_2) + E(A_1 B_2) - 2 E(A_1 C_1) + 2 E(A_2 C_1) - 2 E(A_2 B_1 C_1) - 2 E(A_2 B_2 C_1) - E(A_1 C_2) + E(A_2 C_2) + E(B_1 C_2) - 2 E(A_1 B_1 C_2) - E(A_2 B_1 C_2) - E(B_2 C_2) + E(A_1 B_2 C_2) + 2 E(A_2 B_2 C_2)$\\\vspace{6pt}} \\ \hline
  33 &  \parbox[t]{5in}{\raggedright $0 \leq 6 - E(A_1) - E(A_2) - E(B_1) + E(A_2 B_1) - E(B_2) + E(A_1 B_2) - E(C_1) + E(A_2 C_1) - 2 E(A_2 B_1 C_1) + E(B_2 C_1) - 2 E(A_1 B_2 C_1) - E(A_2 B_2 C_1) - E(C_2) + E(A_1 C_2) + E(B_1 C_2) - 2 E(A_1 B_1 C_2) - E(A_2 B_1 C_2) - E(A_1 B_2 C_2) + 3 E(A_2 B_2 C_2)$\\\vspace{6pt}} \\ \hline
  34 &  \parbox[t]{5in}{\raggedright $0 \leq 6 - E(A_1) - E(A_2) - E(B_1) + E(A_2 B_1) - E(B_2) + E(A_1 B_2) - E(C_1) + E(A_2 C_1) + E(B_1 C_1) + 2 E(A_1 B_1 C_1) - E(A_2 B_1 C_1) + 2 E(B_2 C_1) - 2 E(A_1 B_2 C_1) - 2 E(A_2 B_2 C_1) - E(C_2) + E(A_1 C_2) + 2 E(B_1 C_2) + E(B_2 C_2) - E(A_1 B_2 C_2) + 2 E(A_2 B_2 C_2)$\\\vspace{6pt}} \\ \hline
  35 &  \parbox[t]{5in}{\raggedright $0 \leq 6 - E(A_1) - E(A_2) - E(B_1) + E(A_1 B_1) + 2 E(A_2 B_1) - E(B_2) + 2 E(A_1 B_2) + E(A_2 B_2) - E(A_1 C_1) + E(A_2 C_1) + E(A_1 B_1 C_1) - E(A_2 B_1 C_1) + 2 E(A_1 B_2 C_1) - 2 E(A_2 B_2 C_1) - E(B_1 C_2) + 2 E(A_1 B_1 C_2) + E(A_2 B_1 C_2) + E(B_2 C_2) - 2 E(A_1 B_2 C_2) - E(A_2 B_2 C_2)$\\\vspace{6pt}} \\ \hline
  36 &  \parbox[t]{5in}{\raggedright $0 \leq 6 - 2 E(A_1) - E(A_1 B_1) - E(A_2 B_1) - E(A_1 B_2) - E(A_2 B_2) - E(A_1 C_1) - E(A_2 C_1) - E(B_1 C_1) + 2 E(A_1 B_1 C_1) - E(A_2 B_1 C_1) + E(B_2 C_1) - E(A_1 B_2 C_1) + 2 E(A_2 B_2 C_1) - E(A_1 C_2) - E(A_2 C_2) + E(B_1 C_2) - E(A_1 B_1 C_2) + 2 E(A_2 B_1 C_2) + E(B_2 C_2) - 2 E(A_1 B_2 C_2) + E(A_2 B_2 C_2)$\\\vspace{6pt}} \\ \hline
  37 &  \parbox[t]{5in}{\raggedright $0 \leq 6 - 2 E(A_1) - E(A_1 B_1) - E(A_2 B_1) - E(A_1 B_2) - E(A_2 B_2) - E(A_1 C_1) - E(A_2 C_1) - E(B_1 C_1) + 3 E(A_1 B_1 C_1) + E(B_2 C_1) - 2 E(A_1 B_2 C_1) + E(A_2 B_2 C_1) - E(A_1 C_2) - E(A_2 C_2) + E(B_1 C_2) - 2 E(A_1 B_1 C_2) + E(A_2 B_1 C_2) + E(B_2 C_2) - E(A_1 B_2 C_2) + 2 E(A_2 B_2 C_2)$\\\vspace{6pt}} \\ \hline
  38 &  \parbox[t]{5in}{\raggedright $0 \leq 6 - 2 E(A_1) - 2 E(A_1 B_1) - 2 E(A_2 B_1) - E(A_1 C_1) - E(A_2 C_1) + E(B_1 C_1) - E(A_1 B_1 C_1) + 2 E(A_2 B_1 C_1) - E(B_2 C_1) + 2 E(A_1 B_2 C_1) - E(A_2 B_2 C_1) - E(A_1 C_2) - E(A_2 C_2) + E(B_1 C_2) - E(A_1 B_1 C_2) + 2 E(A_2 B_1 C_2) + E(B_2 C_2) - 2 E(A_1 B_2 C_2) + E(A_2 B_2 C_2)$\\\vspace{6pt}} \\ \hline
  39 &  \parbox[t]{5in}{\raggedright $0 \leq 6 - 2 E(A_1) - 2 E(B_1) + E(A_1 B_1) - E(A_2 B_1) - E(A_1 B_2) - E(A_2 B_2) - 2 E(C_1) + E(A_1 C_1) - E(A_2 C_1) + E(B_1 C_1) - 2 E(A_1 B_1 C_1) + E(A_2 B_1 C_1) - E(B_2 C_1) + E(A_1 B_2 C_1) + 2 E(A_2 B_2 C_1) - E(A_1 C_2) - E(A_2 C_2) - E(B_1 C_2) + E(A_1 B_1 C_2) + 2 E(A_2 B_1 C_2) - E(B_2 C_2) + 2 E(A_1 B_2 C_2) - E(A_2 B_2 C_2)$\\\vspace{6pt}} \\ \hline
  40 &  \parbox[t]{5in}{\raggedright $0 \leq 6 - 2 E(A_1) - 2 E(A_2) - 2 E(B_1) + E(A_1 B_1) + E(A_2 B_1) - E(A_1 B_2) - E(A_2 B_2) - E(A_1 C_1) - E(A_2 C_1) - 2 E(B_1 C_1) + E(A_1 B_1 C_1) + E(A_2 B_1 C_1) - 2 E(B_2 C_1) + 2 E(A_1 B_2 C_1) + 2 E(A_2 B_2 C_1) - E(A_1 C_2) + E(A_2 C_2) + 2 E(A_1 B_1 C_2) - 2 E(A_2 B_1 C_2) - E(A_1 B_2 C_2) + E(A_2 B_2 C_2)$\\\vspace{6pt}} \\ \hline
  41 &  \parbox[t]{5in}{\raggedright $0 \leq 7 - E(A_1) - E(B_1) - E(A_1 B_1) - E(C_1) - E(A_2 C_1) + 3 E(A_1 B_1 C_1) + E(A_2 B_1 C_1) - E(B_2 C_1) + E(A_1 B_2 C_1) + 2 E(A_2 B_2 C_1) - E(A_1 C_2) + E(A_2 C_2) - E(B_1 C_2) + 4 E(A_1 B_1 C_2) - E(A_2 B_1 C_2) + E(B_2 C_2) - E(A_1 B_2 C_2) - 2 E(A_2 B_2 C_2)$\\\vspace{6pt}} \\ \hline
  42 &  \parbox[t]{5in}{\raggedright $0 \leq 8 - E(A_1) - E(A_2) - E(B_1) - E(A_1 B_1) - E(B_2) + E(A_2 B_2) - E(A_1 C_1) + E(A_2 C_1) - E(B_1 C_1) + 2 E(A_1 B_1 C_1) + E(A_2 B_1 C_1) + E(B_2 C_1) + E(A_1 B_2 C_1) - 4 E(A_2 B_2 C_1) - 2 E(A_2 C_2) + E(A_1 B_1 C_2) + 3 E(A_2 B_1 C_2) - 2 E(B_2 C_2) + 3 E(A_1 B_2 C_2) + E(A_2 B_2 C_2)$\\\vspace{6pt}} \\ \hline
  43 &  \parbox[t]{5in}{\raggedright $0 \leq 8 - 2 E(A_1) - 2 E(B_1) + E(A_1 B_1) - E(A_2 B_1) - E(A_1 B_2) + E(A_2 B_2) - E(A_1 C_1) - E(A_2 C_1) - E(B_1 C_1) + 2 E(A_1 B_1 C_1) + 3 E(A_2 B_1 C_1) + E(B_2 C_1) - E(A_1 B_2 C_1) - 2 E(A_2 B_2 C_1) - E(A_1 C_2) + E(A_2 C_2) - E(B_1 C_2) + 3 E(A_1 B_1 C_2) - E(B_2 C_2) + 4 E(A_1 B_2 C_2) - E(A_2 B_2 C_2)$\\\vspace{6pt}} \\ \hline
  44 &  \parbox[t]{5in}{\raggedright $0 \leq 8 - 2 E(A_1) - 2 E(A_2) - 2 E(A_1 B_1) + 2 E(A_2 B_1) - E(A_1 C_1) + E(A_2 C_1) + 2 E(B_1 C_1) - 2 E(A_1 B_1 C_1) - 2 E(A_2 B_1 C_1) - 2 E(B_2 C_1) + E(A_1 B_2 C_1) + 3 E(A_2 B_2 C_1) - E(A_1 C_2) + E(A_2 C_2) + 2 E(B_1 C_2) - 2 E(A_1 B_1 C_2) - 2 E(A_2 B_1 C_2) + 2 E(B_2 C_2) - 3 E(A_1 B_2 C_2) - E(A_2 B_2 C_2)$\\\vspace{6pt}} \\ \hline
  45 &  \parbox[t]{5in}{\raggedright $0 \leq 8 - 3 E(A_1) - E(A_2) - 2 E(A_1 B_1) + 2 E(A_2 B_1) - E(A_1 B_2) + E(A_2 B_2) - 2 E(A_1 C_1) + 2 E(A_2 C_1) + 2 E(B_1 C_1) - 2 E(A_1 B_1 C_1) - 2 E(A_2 B_1 C_1) + 2 E(B_2 C_1) - 2 E(A_1 B_2 C_1) - 2 E(A_2 B_2 C_1) - E(A_1 C_2) + E(A_2 C_2) + 2 E(B_1 C_2) - 2 E(A_1 B_1 C_2) - 2 E(A_2 B_1 C_2) - 2 E(B_2 C_2) + 3 E(A_1 B_2 C_2) + E(A_2 B_2 C_2)$\\\vspace{6pt}} \\ \hline
  46 &  \parbox[t]{5in}{\raggedright $0 \leq 10 - 3 E(A_1) - E(A_2) - 3 E(B_1) + 2 E(A_1 B_1) + E(A_2 B_1) - E(B_2) + E(A_1 B_2) + 2 E(A_2 B_2) - 2 E(A_1 C_1) + 2 E(A_2 C_1) - E(B_1 C_1) + 3 E(A_1 B_1 C_1) - 4 E(A_2 B_1 C_1) - E(B_2 C_1) + E(A_1 B_2 C_1) - 2 E(A_2 B_2 C_1) - E(A_1 C_2) - E(A_2 C_2) - 2 E(B_1 C_2) + 3 E(A_1 B_1 C_2) + E(A_2 B_1 C_2) + 2 E(B_2 C_2) - 4 E(A_1 B_2 C_2) - 2 E(A_2 B_2 C_2)$\\\vspace{6pt}} \\ \hline
\end{longtable}

\end{document}